\documentclass[sigconf]{acmart}
\usepackage{dsfont}
\usepackage{xcolor}
\usepackage[ruled,vlined]{algorithm2e}
\usepackage{amsthm}

\usepackage[utf8]{inputenc}
\usepackage[english]{babel}
\usepackage{mathtools}
\usepackage{multicol}
\usepackage{dblfloatfix}

\newcommand{\rx}[1]{{#1}}
\newcommand{\rxx}[1]{{#1}}

\AtBeginDocument{%
  \providecommand\BibTeX{{%
    \normalfont B\kern-0.5em{\scshape i\kern-0.25em b}\kern-0.8em\TeX}}}

\setcopyright{rightsretained}
\copyrightyear{2021}
\acmYear{2021}
\acmDOI{10.1145/1122445.1122456}

\acmConference[OARS workshop at KDD '21]{OARS workshop at SIGKDD 2021: ACM Conference on Knowledge Discovery and Data Mining}{August 14--18, 2021}{Singapore}
\acmBooktitle{OARS-KDD '21: ACM SIGKDD Conference on Knowledge Discovery and Data Mining,
  August 14--18, 2021, Singapore}
\acmPrice{15.00}
\acmISBN{978-1-4503-XXXX-X/18/06}

\begin{document}

\title{Adaptively Optimize Content Recommendation Using Multi Armed Bandit Algorithms in E-commerce}

\author{Ding Xiang}
\affiliation{%
  \institution{The Home Depot}
  \city{Atlanta}
  \country{USA}
}
\email{Ding_Xiang@homedepot.com}

\author{Becky West}
\affiliation{%
  \institution{The Home Depot}
  \city{Atlanta}
  \country{USA}
}
\email{Rebecca_West@homedepot.com}

\author{Jiaqi Wang}
\affiliation{%
  \institution{The Home Depot}
  \city{Atlanta}
  \country{USA}
}
\email{Jiaqi_Wang@homedepot.com}

\author{Xiquan Cui}
\affiliation{%
 \institution{The Home Depot}
  \city{Atlanta}
  \country{USA}
}
\email{Xiquan_Cui@homedepot.com}

\author{Jinzhou Huang}
\affiliation{%
  \institution{The Home Depot}
  \city{Atlanta}
  \country{USA}
}
\email{Jinzhou_Huang@homedepot.com}


\begin{abstract}
\rx{E-commerce sites strive to provide users the most timely relevant} information in order to reduce shopping frictions and \rxx{increase customer satisfaction}.
\rxx{Multi armed bandit models (MAB) as a type of adaptive optimization algorithms provide possible approaches for such purposes.}
\rx{In this paper, we \rxx{analyze using} three classic MAB algorithms, $\epsilon$-greedy, Thompson sampling (TS), and upper confidence bound 1 (UCB1) for dynamic content recommendations, and walk through the process of developing these algorithms internally to solve a real world e-commerce use case.} 
First, we analyze the three MAB algorithms using simulated purchasing datasets with non-stationary reward distributions \rxx{to simulate the possible time-varying customer preferences}, where the traffic allocation dynamics and the accumulative rewards of different algorithms are compared. We find all three algorithms can adaptively optimize the recommendations, and under this simulated scenario UCB1 surprisingly slightly outperforms the most popular TS algorithm. Second, we compare the accumulative rewards of the three MAB algorithms 
with more than 1,000 trials using \rxx{actual historical} A/B test datasets. 
We find that the larger difference \rx{between the success rates} of competing recommendations the more accumulative rewards the MAB algorithms can achieve. In addition, we find that TS shows the highest average accumulative rewards under different testing scenarios. 
Third, we develop a batch-updated MAB algorithm to overcome the delayed reward issue in e-commerce and \rx{enable an online content optimization on our} App homepage. For a state-of-the-art comparison, a \rxx{real} A/B test among our batch-updated MAB algorithm, a third-party MAB solution, and the default business logic are conducted. The result shows that our batch-updated MAB algorithm outperforms the counterparts and achieves 6.13\% relative click-through rate (CTR) increase and 16.1\% relative conversion rate (CVR) increase compared to the default experience, and 2.9\% relative CTR increase and 1.4\% relative CVR increase compared to the external MAB service.

\end{abstract}




\begin{teaserfigure}
\end{teaserfigure}

\maketitle

\section{Introduction}
E-commerce offers customer convenience and large assortment \rx{options} compared to \rx{physical stores}.
But it also creates unique challenges, e.g. how to surface the most relevant information to a customer among massive contents on a limited 2D screen? The most used approach today in e-commerce is to perform an A/B test and pick the winner among a few hypothesized design choices. However, in many cases, a sensible goal might not be to choose a \rx{fixed} winner among competing experiences. For example, if the underlying composition of user population or intent changes dramatically with time, there might not exist a definite winner design for all situations. In addition, under some circumstances, the \rx{opportunity} cost of assigning \rx{some} users to an inferior experience \rx{might be very high and often not reported.}
The constant traffic allocation framework required by an A/B test seems too rigid. To deal with these problems, recently more and more internet companies start using a continuous optimization framework, multi armed bandit (MAB), to maximize the relevancy of their content recommendation dynamically. 

MAB is a type of algorithm that belongs to the reinforcement learning category, or a simple version of reinforcement learning without state transition. It was first posed by researcher Thompson \cite{10.2307/2332286} as a concept of clinical trial design, and later on studied by Robbins \cite{bams/1183517370} and Bellman \cite{10.2307/25048278} in a more general format referred to as sequential design of experiments. The name MAB first appeared in the publications during 1950s such as Neyman \cite{neyman1951berkeley} and Bush et al \cite{bush1953stochastic}. It describes a gambling situation, where in a gambling room there are multiple slot machines, and each machine has its own success distribution. The gambler has to decide by trying and observing which arm of the slot machine to pull, in order to maximize the total money received. 

MAB algorithms are extensively studied \cite{berry1985bandit, slivkins2019introduction, lai1985asymptotically, agrawal1995sample, auer2002finite} in a wide range of applications \cite{huo2017risk, villar2015multi, schwartz2017customer, yu2015large, sutton1998introduction}. In this paper, we mainly focus on three classical multi armed bandit algorithms, $\epsilon$-greedy \cite{sutton1998introduction}, Thompson sampling \cite{russo2017tutorial}, and upper confidence bound 1 (UCB1) \cite{auer2002finite}. More details of these algorithms are given later as we introduce the problem formulations. We analyze these MAB algorithms for content recommendation in the e-commerce settings, where we may face some different issues than what the original MAB algorithms are designed for or the assumptions they use. 

For example, in a typical MAB algorithm, it is assumed that the success distribution of each slot machine, or an arm, is fixed. In the e-commerce setting using MAB for content recommendations implies the slot machine's response is an analogy for customers' feedback. However, oftentimes customers' feedback is non-stationary as their preferences may change over time.
In addition, after pulling an arm of a slot machine, the response is usually available instantaneously. However, this may not be the case in e-commerce, where the customers may take minutes, hours, or even longer to provide feedback, such as a purchase. 

In recent years, some studies provide theoretical analysis of MAB performance in terms of regrets under mathematically tractable non-stationary reward distributions \cite{besbes2014stochastic, slivkins2008adapting, garivier2008upper, cao2019nearly} and some others proposed different algorithms that can handle the delayed issue either theoretically \cite{joulani2013online, guha2010multiarmed} or practically \cite{grover2018best, liu2019multi}. In this paper, we do not focus on the theoretical analysis part. Instead, we analyze the performance and properties of using the MAB algorithms first on simulated non-stationary distributed user purchasing datasets, and then we evaluate the MAB algorithms offline 
using datasets logged in real-world A/B tests of a large e-commerce site. Lastly, we propose a batch-update MAB framework to tackle some of the practical data delay issues, and provide a real online A/B test performance of adaptively optimizing the sequence of content cards on the homepage of a major e-commerce App.

\rxx{The rest of the paper is organized as follows. Section 2 introduces the three MAB algorithms to be studied in this paper. Section 3 analyzes the performance of the three MAB algorithms in the simulated non-stationary scenarios. Section 4 provides the offline evaluation of the three algorithms using actual historical A/B testing datasets. Section 5 illustrates our batch-updated MAB framework with a real three-way online A/B test. Section 6 concludes the paper and provides certain future directions.}

\section{Multi Armed Bandit Algorithms}

Consider we have $K$ competing experiences (i.e., arms), denoted by set $E=\{1, 2, ..., K\}$, and a decision strategy $\boldsymbol S$ such that for every customer's visit at time $t = 1, 2,..., T$, the strategy $\boldsymbol S$ can decide which one of the experiences, $e_t \in E$, to show. After showing the experience $e_t$, we will see a feedback or reward, denoted by $r_t$, from the customer who received the experience. The feedback could either be binary ($r_t \in \{0, 1\}$) such as the experience being click or not, and a purchase being made or not, or continuous ($r_t \in \mathbb{R}, r_t \geq 0$) such as the total price of the order etc.

An MAB problem is described that assuming at each visit time $t$, for any experience $e \in E$ being shown, i.e., $e_t = e$, the corresponding rewards $r_t$ are drawn independently from an unknown distribution $\mathcal{D}_e$, then how we can decide the experience to show at each time in order to maximize the total rewards $\sum_{t=1}^T r_t$. A main challenge in this problem is that for the experiences that are not shown before, their reward distribution is unknown. Hence a common exploration-exploitation dilemma arises, i.e., we want to explore the unknown distributions hoping to find a better experience with higher rewards, but this is on the cost of giving up exploiting the current best experience we have learned so far.

To deal with this problem, many strategies or the MAB algorithms, are proposed and analyzed. In this paper, we mainly focus on the three MAB algorithms, i.e., $\epsilon$-greedy, Thompson sampling, and upper confidence bound 1 (UCB1). 

1) \textbf{$\epsilon$-greedy}: the $\epsilon$-greedy algorithm \cite{sutton1998introduction} is a strategy which assigns at each visit time a small probability $\epsilon$ for exploration (i.e., randomly selecting an experience to show) and with the $1- \epsilon$ probability to exploit the current winner experience (i.e., showing the experience that currently generates the highest average reward based on what it has learned so far). 

2) \textbf{Thompson Sampling (TS)}: different from the $\epsilon$-greedy using a fixed exploration rate, Thompson sampling \cite{russo2017tutorial} keeps adjusting exploration rate based on its current estimation on the reward distributions $\mathcal{D}_e, \forall e \in E$. The approach usually starts from assigning an initialized beta distribution $\boldsymbol{B}(\alpha_e, \beta_e > 1)$ to each experience $e \in E$, which is used to gradually approximate the true average reward distribution of the experience. At every visit time, the model generates a sample from the currently learned beta distributions of all experiences, and the experience with the largest sample is shown to the customer. Then based on the customer's feedback, the model updates the beta distribution related to that shown experience. For binary rewards $r_{t}$, this update can be very efficient, i.e., $(\alpha_{e_{t+1}}, \beta_{e_{t+1}}) \leftarrow(\alpha_{e_{t}} + r_{t}$,  $\beta_{e_{t}} - r_{t} + 1)$. (the beta distributions of unselected experiences remain the same.)

3) \textbf{Upper Confidence Bound 1 (UCB1)}: This algorithm balances the exploration and exploitation by using upper confidence bounds of the current average rewards for each experience \cite{auer2002finite}. At each visit time the experience with the highest upper confidence bound is shown to the customer. An upper confidence bound of an experience $e$ at time $t$ consists of two parts, the current average reward of the experience $p_e(t)$, and an upper confidence range (with high probability) $\sqrt{\frac{2\ln{(t )} }{N_{t}(e) + 1}}$, where $N_t(e)$ is the number of times experience $e$ is shown by time $t$.

The theoretical analysis of these three algorithms in terms of their expected total rewards, or usually equivalently measured by the expected regrets, and their asymptotic relationship with time steps, can be found in \cite{slivkins2019introduction}. In this paper we mainly focus on the application and performance analysis of these algorithms.

\section{Performance of MAB in simulated non-stationary scenarios}
\subsection{Simulated Datasets and Methodology}
\label{datasets}
\begin{figure}[h]
  \centering
  \includegraphics[width=\linewidth]{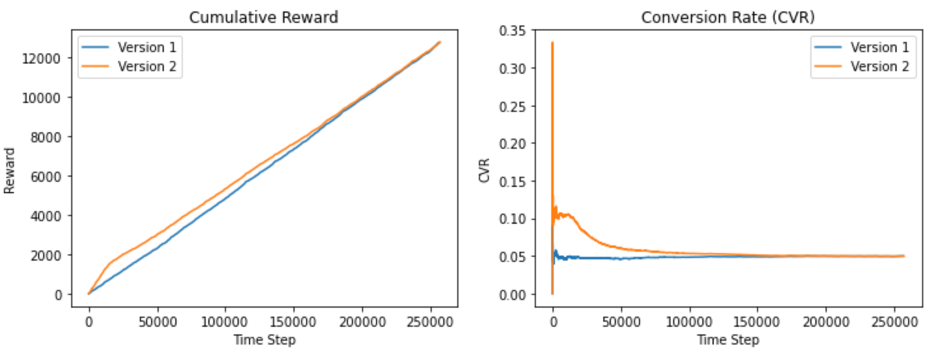}
  \vspace{-0.6cm}
  \caption{Accumulated purchases}
  \label{fig-name1}
\end{figure}

In order to illustrate how the different MAB algorithms work under a non-stationary scenario while protecting data privacy, 
we created a simulated dataset to represent a common and challenging practical situation where one or more distributions are non-stationary. Specifically, we added some disturbance. One experience (v2) has a higher local conversion rate (CVR) than the other (v1) at the beginning of the test, but a lower local CVR \rxx{(than v1)} later on, so that the two experiences finally converge to the same CVR (average conversion over the whole time range). The cumulative reward (purchases) and CVR of the simulated dataset are presented in Figure \ref{fig-name1}.


\rx{Since the simulated datasets are generated with a uniformly random logging policy, i.e., at each time step a version is chosen uniformly, we can use the unbiased offline evaluation method proposed by Li \textit{et. al} \cite{li2011unbiased}, to test the performance of the three MAB algorithms and the A/B test sampling strategy with the simulated dataset.}

\subsection{Traffic Allocation Patterns}
\begin{figure}[h]
  \centering
  \captionsetup{justification=centering}
  \includegraphics[width=\linewidth]{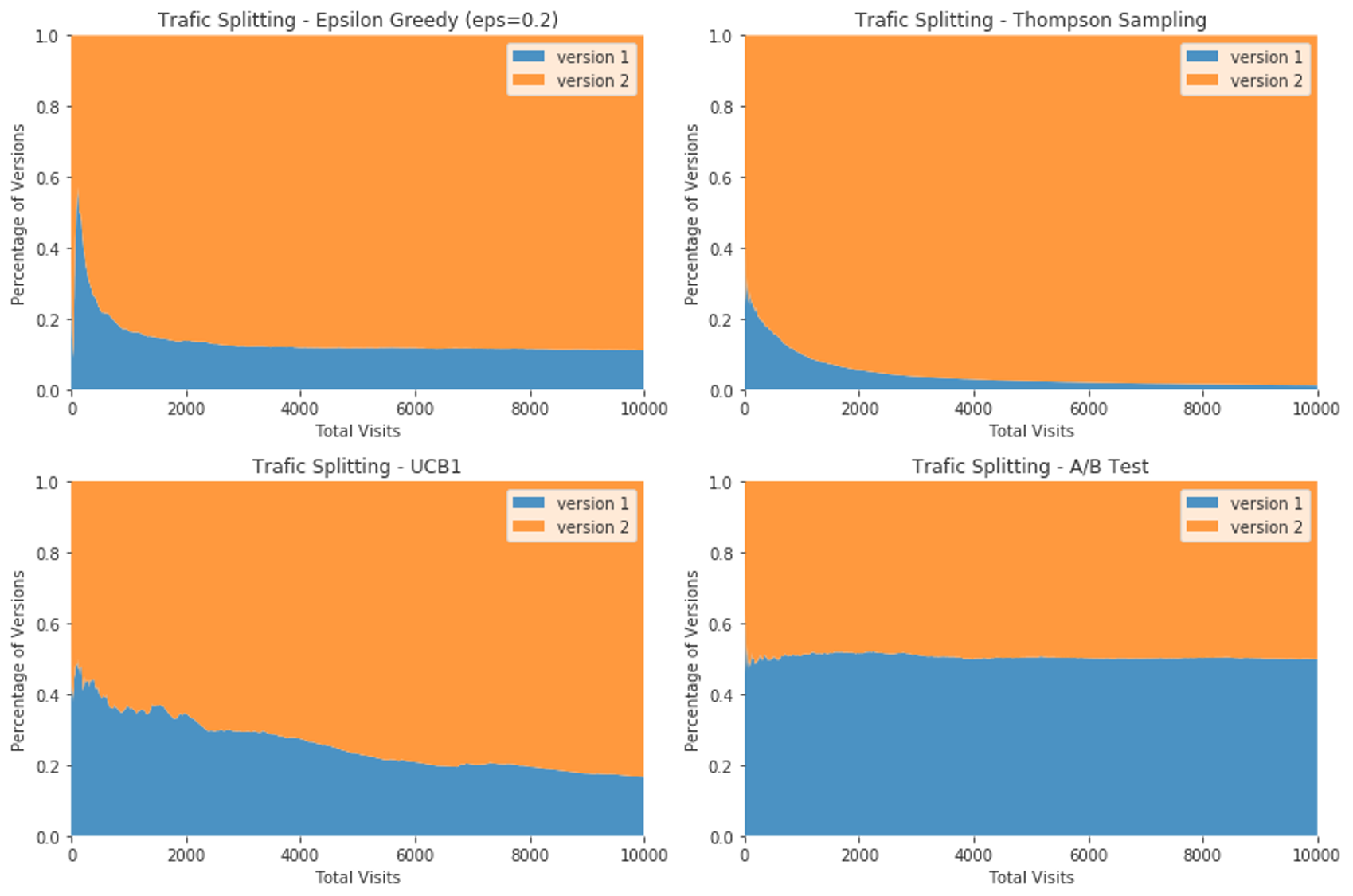}
  \caption{Traffic allocation patterns of the MAB algorithms\\ and typical A/B testing}
  \label{fig-name3}
  \vspace{-0.3cm}
\end{figure}

We first analyzed the traffic allocation dynamics for these 4 different strategies, which are shown in Fig. \ref{fig-name3}. In the figure, we see that compared to the A/B test strategy all the three MAB algorithms can 
adjust the traffic allocation earlier, 
i.e., gradually increasing the traffic amount for the winner experience (v2) and lowering the traffic amount for the under-performed experience (v1).

For the tested three MAB algorithms, $\epsilon$-greedy algorithm ($\epsilon = 0.2$) shows the highest traffic changing speed at the beginning. However, after
detecting a relatively stable winner, 
its traffic allocation for the winner is maintained around the $90\% = (1-0.2) + 0.2/2$ level. Thompson sampling not only shows a high adjustment speed at the beginning, but also continuously shows an aggressive allocation of traffic towards the winning experience. It is the only one that keeps the currently detected winner (version 1) at a very dominate level (around $98\%$ of the total traffic) while keeps the losing version traffic only around $2\%$.
The traffic allocation dynamics for UCB1 seems more "moderate", i.e., the changing speed is relatively lower than the other two MAB algorithms.

\subsection{Reward and Empirical Regret Comparison}
\begin{figure}[h]
\captionsetup{justification=centering}
  \begin{minipage}{0.235\textwidth}
  \includegraphics[width=\linewidth]{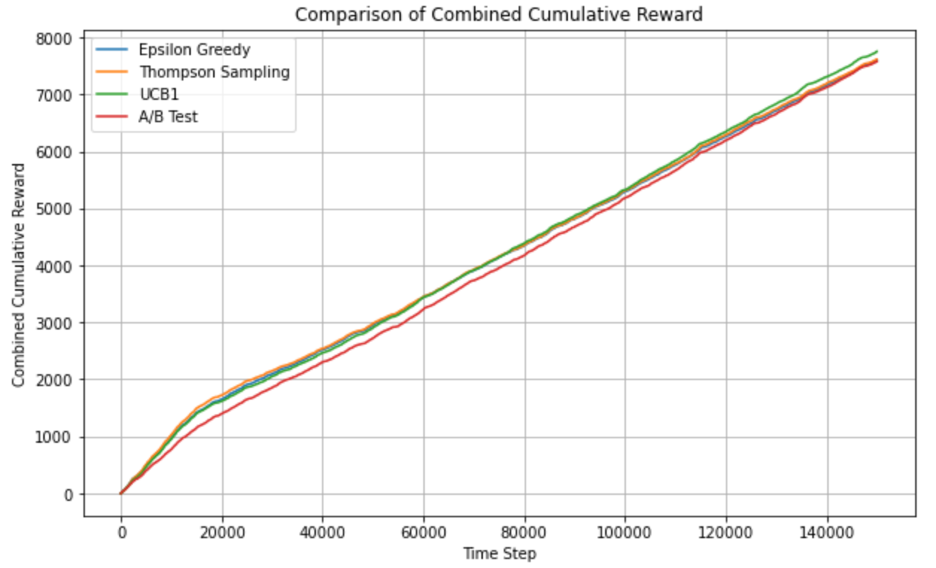}
  \caption{Total rewards}
  \label{total_rewards}
\end{minipage}
\hfill
\begin{minipage}{0.235\textwidth}
  \includegraphics[width=\linewidth]{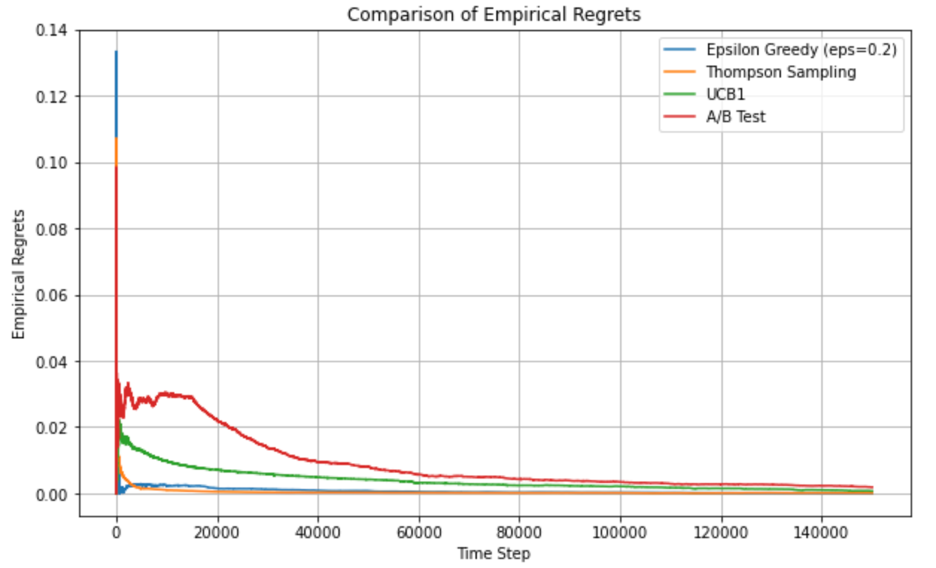}
  \caption{Empirical regrets}
  \label{regrets}
\end{minipage}
\vspace{-0.5cm}
\end{figure}

\begin{figure}[h]
\captionsetup{justification=centering}
  \begin{minipage}{0.235\textwidth}
  \includegraphics[width=\linewidth]{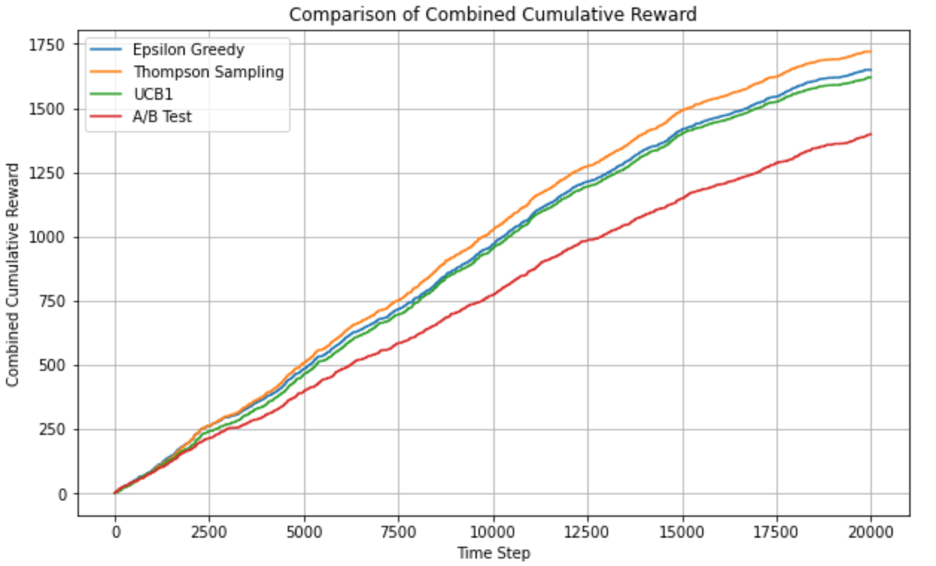}
  \caption{Total rewards \\early-half zoom-in}
  \label{total_rewards_1}
\end{minipage}
\hfill
\begin{minipage}{0.235\textwidth}
  \includegraphics[width=\linewidth]{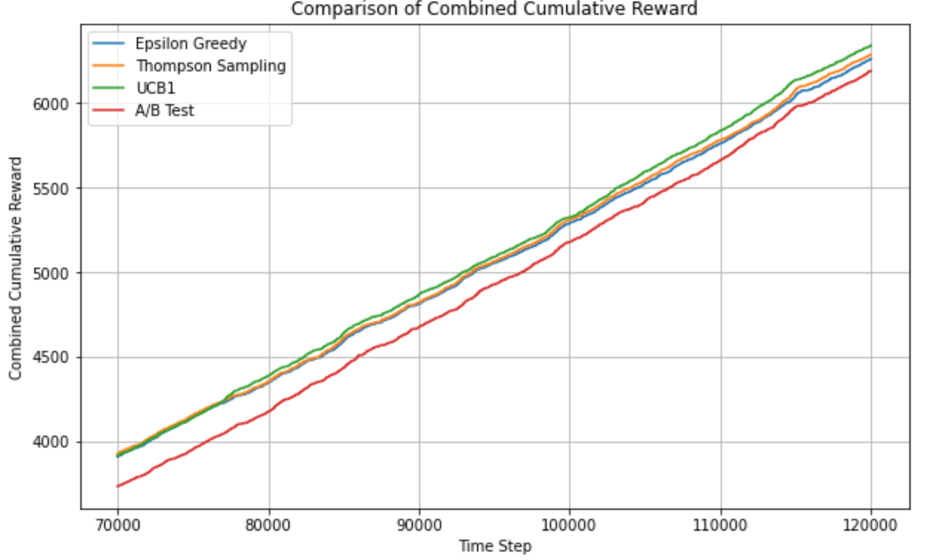}
  \caption{Total rewards\\ later-half zoom-in}
  \label{total_rewards_2}
\end{minipage}
\end{figure}

The total rewards for different algorithms over the dynamic recommendations are shown in Fig. \ref{total_rewards}, where we see Thompson sampling achieves a higher total rewards than the others in the early half, as shown in the zoom-in plot Fig. \ref{total_rewards_1}, while interestingly UCB1 achieves the highest total rewards in the later half thus outperforms TS as displayed in the zoom-in plot Fig \ref{total_rewards_2}. This result is interesting as Thompson sampling recently has gain wide popularity because of its strong performance as demonstrated both empirically and analytically, which seems to suggest that TS would always be the top choice. However, this simulation generates an opposite result.
We would like to highlight that the result here does not mean in general or more practical cases TS is definitely worse than UCB1 (in fact we provide a more comprehensive comparison in the later section under industrial datasets that TS does show a better performance). This simulation result can be regarded as an "adversarial" case, where the fast-and-aggressive traffic adjustment pattern of TS, happens to be a weaknesses here, since TS does not detect the higher local CVR of version 1 in the later half and still assigns the dominate traffic to version 2 according to the whole time CVR. This implies that no MAB algorithm can always be the "perfect" algorithm; thus the choice should depend on the real use cases.

We also compare the empirical average regrets of the three different algorithms, as shown in figure \ref{regrets}, since "regrets" are also commonly used metrics for evaluating MAB algorithms. However, different from the theoretical average regrets at a given time $T$, i.e.,
\[
    g(T) = \frac{T\mu^* - \sum_{t=1}^T r_{j(t)}}{T},
\]
where $\mu^*$ is the true success rate of the winner experience,
here we use empirical regrets, defined by, 
\[ 
    \hat{g}(T) = \frac{T\hat{\mu}(T)^* - \sum_{t=1}^T r_{j(t)}}{T},
\] 
where $\hat{\mu}(T)^*$ is the empirical success rate of the winner experience based on what the model has learned up to time $T$,  
to simulate the real situations where we have no prior knowledge of the reward distribution. (thus the regret can only be estimated based on the experience). From the figure \ref{regrets}, we can see that UCB1 has a relatively higher empirical regret per trial, which seems "opposite" to its good performance in terms of the total rewards obtained. This is because the empirical regrets are purely determined by the best distribution at the current step the agent has learned so far and each agent learns in a different way. 

To illustrate how higher rewards do not always correlate with lower empirical regrets, we show a toy example in Table~\ref{tab:regretreward}. In this example, two different arms are tested against each other using two different algorithms. In this case, we assume conversion rate for Arm 2 starts off lower then increases at a later stage during the test. Algorithm 1 more aggressively shifts traffic to Arm 1 at the beginning of the test while Algorithm 2 adapts to this change and gives a more balanced traffic allocation. In this example, both algorithms achieve the same regret but Algorithm 1 has higher rewards.


\begin{table}[h]
\captionsetup{skip=6pt, justification=centering}
\caption{The algorithm with the lowest empirical regret may not have the largest reward.}
\label{tab:regretreward}
\begin{tabular}{|c c|c|c|}

\hline
 \rule{0pt}{10pt} &  & Algorithm 1 & Algorithm 2 \\[1.5pt]
\hline
 \rule{0pt}{10pt} & Trials & 800 & 500 \\
Arm 1 & Wins & 400 & 240 \\
 & CTR & 0.5 & 0.48 \\[2pt]
\hline
 \rule{0pt}{10pt} & Trials & 200 & 500 \\
Arm 2 & Wins & 20 & 160 \\
 & CTR & 0.1 & 0.32 \\[2pt]
 \hline
 \multicolumn{2}{|c|}{\rule{0pt}{10pt}{Empirical Regret}} & 80 & 80 \\
 \multicolumn{2}{|c|}{Reward}  & \textbf{420} & \textbf{400} \\[2pt]

\hline
\end{tabular}
\end{table}

\section{Offline Evaluation of MAB algorithms on Industrial Datasets}

\begin{figure}[h]
  \centering
  \captionsetup{justification=centering}
  \includegraphics[width=0.95\linewidth]{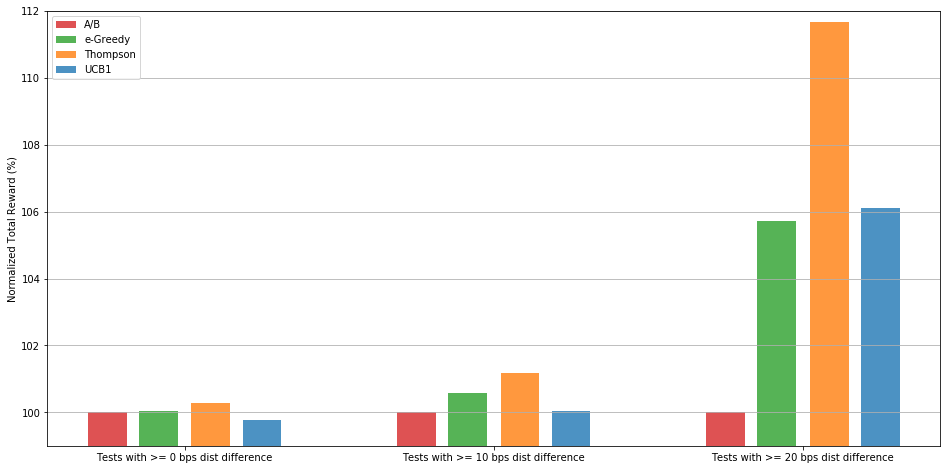}
  \caption{Comparison of normalized performance of different MAB algorithms under different scenarios}
  \label{hist_ab}
\end{figure}

In order to estimate the performance of using each MAB algorithm in industrial settings, we evaluate their performance on the historical A/B testing datasets of a major e-commerce website based on the unbiased offline evaluation method \cite{li2011unbiased} with around 1000 offline trials, where all test durations are longer than 2 weeks and the visits are in the order of \rxx{10 thousands or millions}. The performance is shown in figure \ref{hist_ab}.  

\begin{figure*}[h]
  \centering
  \includegraphics[width=0.95\linewidth]{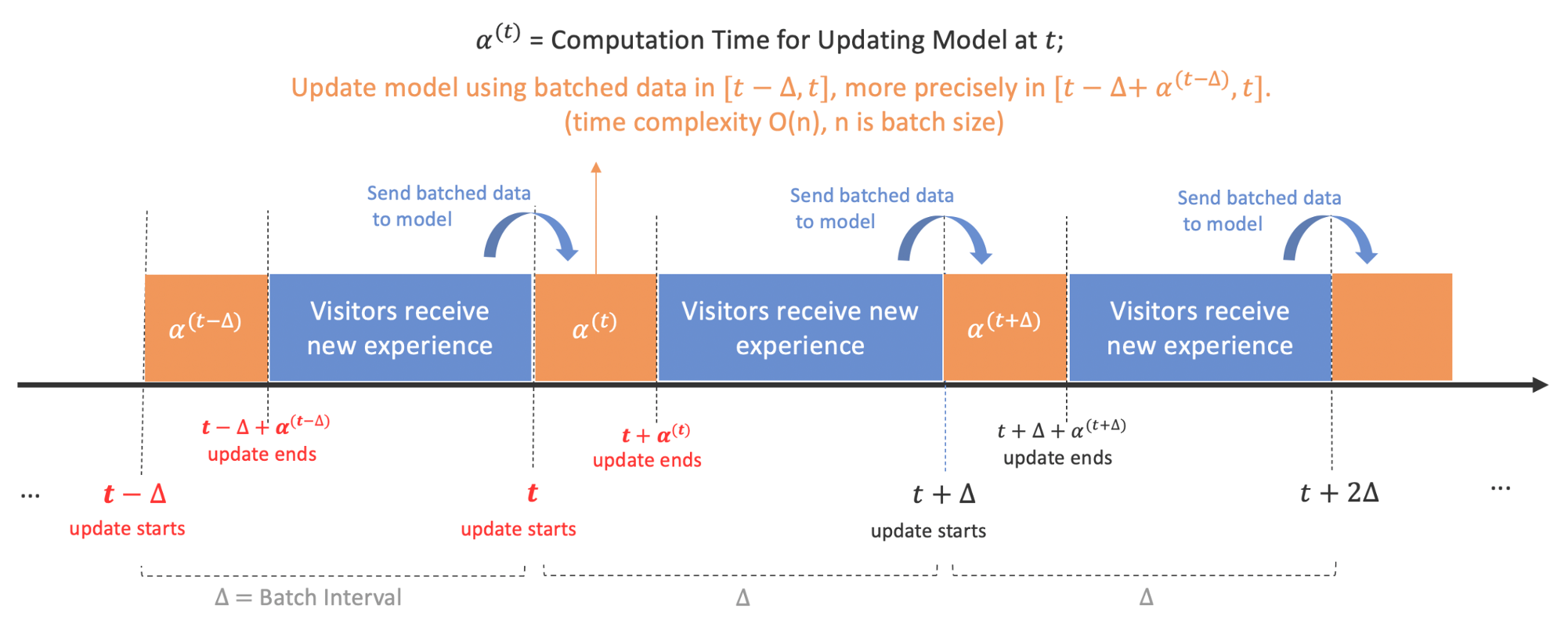}
  \caption{Unrolled Illustration of Batch Update MAB Framework}
  \label{batch_MAB}
\end{figure*}

In the above figure, there are three types of testing scenarios, where the first one on the left area is for the performance of the tests where the true success rate difference (computed based on the whole datasets) between the control and treatment groups are larger than or equal to 0 basis point (BPS) 
, while the tests in the middle are for the true success rate difference \rxx{larger than or equal to} 10 BPS, and the last tests on the right are for the success rate difference \rxx{$\geq$} 20 BPS. The y axis describes the relative total reward. To protect confidential information, we normalize the average reward for A/B testing to $1$ as a general benchmark. It can be seen that as the true success rate difference (or "gap") increases, the performance of the MAB algorithms also improves. Also we noticed that Thompson sampling shows the highest average reward improvements in all three scenarios. Thus it has the strongest performance based on the offline evaluation. Epsilon greedy algorithm also performs better than A/B test in the three scenarios (thus also robust), even though it does not show the best performance as Thompson sampling does. UCB1 on the other hand, performs the worst in the first scenario, but as the "gap" increases it outperforms A/B tests or even epsilon greedy under the largest difference scenario. 

\section{Online Optimization Using a Batch-Update MAB Framework}

\begin{figure}[b]
  \centering
  \captionsetup{justification=centering}
  \includegraphics[width=0.95\linewidth]{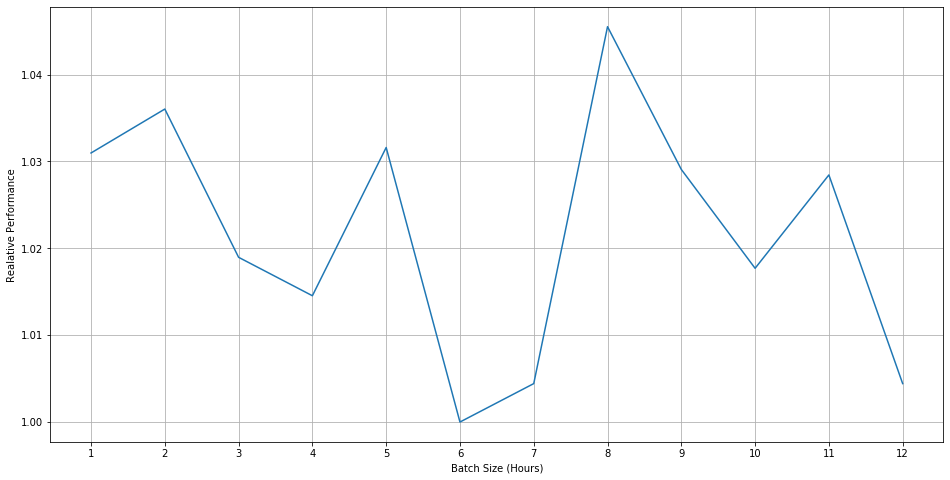}
  \caption{Performance comparison under different batch sizes for $\epsilon$-greedy algorithms ($\epsilon = 0.2$)}
  \label{batch_size}
\end{figure}

As mentioned before in some e-commerce use cases with possible time-varying customer preferences and considerable opportunity cost, a continuous optimization of the content recommendation is probably a better choice than deciding a fixed winner. Hence MAB algorithms provide possible solutions in these use cases. However, one of the big challenges to implement MAB algorithm for content recommendation in a large e-commerce platform like in The Home Depot is to deal with the data latency issue. The data latency issue is two-folds: 1) data refresh delay due to existed data platform configuration. 2) reward data delay due to customers' late feedback for the experience they receive. To solve the issue, we design a batch-update MAB framework, which is shown in figure \ref{batch_MAB}. The lower bound of the batch update should be the data fresh latency. With the lower bound, a practical and good choice of batch size can be decided by offline evaluations on the historical datasets that are related to the business cases under consideration. As an example, figure \ref{batch_size} shows the performance of using different batch-size with $\epsilon$-greedy algorithms ($\epsilon = 0.2$), where in this case 8-hour batch size shows the best performance.

\begin{figure}[h]
  \centering
  \captionsetup{justification=centering}
  \includegraphics[width=\linewidth]{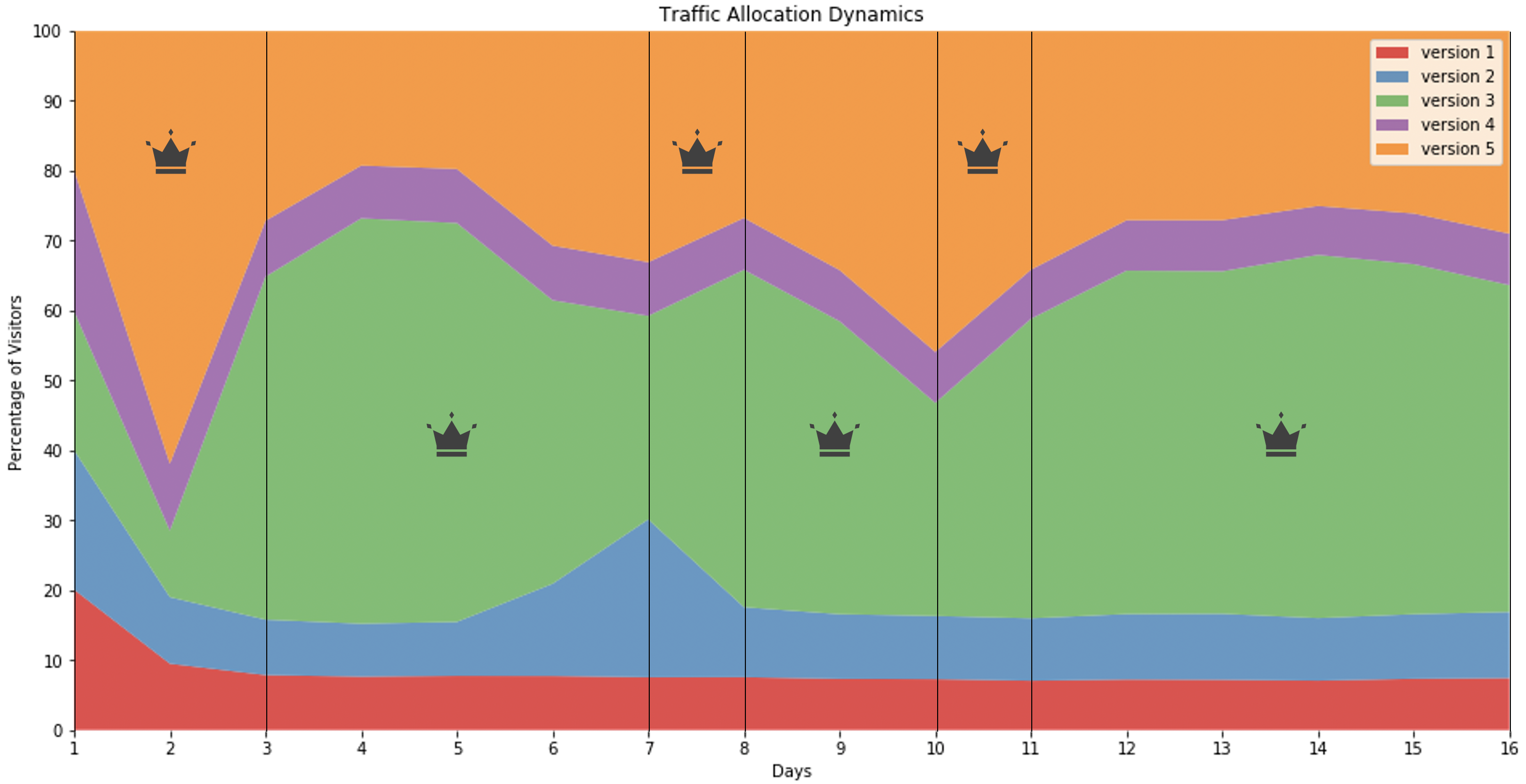}
  \caption{Traffic allocation and detected winner experience of the batch MAB algorithm in the A/B test}
  \label{AB_test}
\end{figure}

To test the performance of the batch-update MAB framework, we apply it on the homepage of our App to optimize the order of content cards (widgets) and maximize its click through rate or conversation rate. On our homepage, there are five different orders of content cards (i.e., five different experiences). 
Our goal is to continuously find the best experience to fit the needs of the user population at the moment. For a state-of-the-art comparison, a real A/B test of our batch-updated MAB algorithm, a third-party MAB solution, and the default business logic were conducted. With millions of visits from real online shoppers who engaged with the App homepage, we saw 6.13\% relative increase in the click-through-rate (CTR) and 16.1\% relative increase in the conversion rate (CVR) compared to the default experience, and 2.9\% relative CTR increase and 1.4\% relative CVR increase compared to the external MAB service. 
Moreover, we do observe that our user base behaves differently on different days of the test, and MAB allocates different proportions of visits to different design choice according to their performance at the time. The traffic dynamics (in percentage) and the detected winner experience are shown in figure \ref{AB_test}.

The details of the test design and metrics values are confidential thus are not disclosed here.

\section{Conclusion}
In this paper, we analyze using the three classic MAB algorithms, $\epsilon$-greedy, Thompson sampling (TS), and upper confidence bound 1 (UCB1) for dynamic content recommendation. Under simulated purchasing datasets with non-stationary reward distributions, we find
\begin{itemize}
    \item All the three MAB algorithms can adaptively adjust traffic and achieve a higher total rewards compared to random sampling in A/B testing. 
    \item TS shows the most aggressive traffic allocations among all three algorithms, while UCB1 shows the most moderate traffic allocations.
    \item TS can be outperformed by UCB1 in the non-stationary reward distribution case (e.g the "adversarial" case here). 
\end{itemize}
Second, in the offline evaluation based on industrial datasets, we find that
\begin{itemize}
    \item The larger difference between the competing experiences in terms of the success rate, the more total rewards the MAB algorithms can achieve. 
    \item TS shows the strongest performance in terms of the total rewards obtained under different offline testing scenarios. 
\end{itemize}
Last, a batch-updated MAB algorithm is proposed to overcome the practical data latency issues and enable the real world content optimization on the homepage of a major e-commerce App. The real A/B test shows our batch-updated MAB algorithm outperformed the counterparts and achieved 6.13\% increase in click-through rate and 16.1\% increase in conversion rate.

\rx{Future directions include designing new MAB algorithms that can achieve higher rewards by considering personalization potentials under the possible time-varying customer preferences and feedback delays, and designing new adaptive optimization algorithms that are compatible with more general business success metrics other than click through rate and conversion rate etc. to increase the flexibility of a content recommendation framework.} 

\section{Acknowledgement}

We thank Kevin Abdo's continuous support allowing us to apply and test our MAB framework in the App homepage use case. We thank Sarfaraz Hussein for providing the testing variants. We thank Priyanka Kommidi and Achal Dalal for the effort and support in the A/B test. We thank Weiyu Li, Varun Vohra, Shradha Mani, Evan Bailey and people who contributed in the data platform development and deployment process.



%
%



\bibliographystyle{ACM-Reference-Format}
\bibliography{sample-base}

\appendix

\end{document}